\documentclass[final,5p,times,twocolumn]{elsarticle}

\usepackage{graphicx,amsmath,amssymb,bm,multirow}
\usepackage{amsthm,mathrsfs}
\usepackage{amsfonts}

\usepackage[usenames,dvipsnames]{color}
\usepackage[colorlinks,linkcolor=blue,anchorcolor=blue,citecolor=blue]{hyperref}

\newcommand{\red}[1]{\textcolor{red}{#1}}
\renewcommand{\vec}[1]{\mbox{\boldmath $#1$}}

\usepackage[nodots]{numcompress}

\begin{document}

\begin{frontmatter}

\title{Microscopic particle-rotor model for low-lying spectrum of  $\Lambda$ hypernuclei}

\author{H. Mei\fnref{1,2}}
\author{K. Hagino\fnref{1,3}}
\author{J.M. Yao\fnref{1,2}}
\author{T. Motoba\fnref{4,5}}
\address[1]{Department of Physics, Tohoku University, Sendai 980-8578,Japan}
\address[2]{School of Physical Science and Technology,
             Southwest University, Chongqing 400715, China}
\address[3]{Research Center for Electron Photon Science, Tohoku University,
            1-2-1 Mikamine, Sendai 982-0826, Japan}
\address[4]{Laboratory of Physics, Osaka Electro-Communications University,
Neyagawa 572-8530, Japan}
\address[5]{Yukawa Institute for Theoretical Physics, Kyoto University, Kyoto 606-8502, Japan }
%
\begin{abstract}
We propose a novel method for low-lying states of hypernuclei based on the particle-rotor
model, in which hypernuclear states are constructed by coupling the hyperon to low-lying states of the core nucleus.
In contrast to the conventional particle-rotor model, we employ a microscopic
approach for the core states, that is, the generator coordinate method (GCM)
with the particle number and angular momentum projections. We apply this microscopic particle-rotor model to $^9_\Lambda$Be employing a
point-coupling version of the relativistic mean-field Lagrangian. A reasonable agreement with the experimental data for the
low-spin spectrum is achieved using the $\Lambda N$  coupling strengths determined
to reproduce the binding energy of the $\Lambda$ particle.
\end{abstract}


\end{frontmatter}

%
%

{\em Introduction}.$-$
In the past decade, many high-resolution $\gamma$-ray spectroscopy experiments have been carried out for $p$-shell $\Lambda$-hypernuclei~\cite{Hashimoto06,Tamura09}. The measured energy spectra and electric multipole transition strengths in the
low-lying states provide rich information on the
$\Lambda$-nucleon interaction in nuclear medium and on the
impurity effect of $\Lambda$ particle on
nuclear structure. In this context, several interesting phenomena have been disclosed. One of the most important findings is
the appreciable shrinkage of the nuclear core due to the $\Lambda$ participation
\cite{Motoba83,Motoba85,Hiyama99}, for which a theoretical
prediction has been clearly confirmed in the experiment \cite{Tanida01}.

The theoretical studies of $\gamma$-ray spectroscopy for $p$-shell
hypernuclei have been mainly performed with
the cluster model~\cite{Motoba83,Hiyama99,Bando90,Hiyama03}
and with the
shell model~\cite{Dalitz78,Gal71,Millener}.
Recently, an ab-initio method as well as
the antisymmetrized molecular dynamics (AMD)
have also been extended
in order to study low-lying states of hypernuclei~\cite{abinitio,Isaka11}.
Most of these models, however, have been limited to light hypernuclei
while it
may be difficult
to apply them to medium-heavy and heavy hypernuclei.

A self-consistent mean-field approach offers a way to
study globally the structure of atomic nuclei as well as hypernuclei from light to heavy
systems \cite{Bender03}, although the pure mean-field approximation does not
yield a spectrum of nuclei due to the broken symmetries.
In the recent decade, the self-consistent mean-field models
have been applied to study the impurity effect of $\Lambda$ particle
on the nuclear
deformation of $p$- and $sd$-shell $\Lambda$ hypernuclei
~\cite{Zhou07,Win08,Schulze10,Win11,Lu11,Li13,Lu14}.
It has been found that
the shape polarization
effect of $\Lambda$ hyperon is in general not prominent,
except for a few exceptions,
including $^{13}_{~\Lambda}$C, $^{23}_{~\Lambda}$C,
and $^{29,31}_{~~~~\Lambda}$Si~\cite{Win08}.

These mean-field studies have shown that the potential
energy surface of a hypernucleus is generally softer against deformation
than that of the corresponding core nucleus.
This implies that the shape fluctuation effect, which is not
included in the pure mean-field approximation,
will be more important in hypernuclei than in normal nuclei.
Furthermore, in order to connect mean-field results to
spectroscopic observables, such as $B(E2)$ values,
one has to
rely on additional assumptions such as the rigid rotor model,
which however would not work for, {\it e.g.,}
nuclei with small deformation or with shape coexistence.
To quantify the impurity effect of $\Lambda$ particle
on nuclear structure,
one thus has to go beyond the pure
mean-field approximation.

Recently, we have quantitatively
studied the impurity effect of $\Lambda$ hyperon
on the low-lying states of $^{24}$Mg
by using a five-dimensional collective Hamiltonian as a choice of the beyond mean-field
approaches~\cite{Yao11NPA}. To this end, we have used
parameters determined by triaxially
deformed Skyrme-Hartree-Fock+BCS calculations.
We have applied
this method to transition strengths and
found that the presence of one $\Lambda$ hyperon
in the $s_{1/2}$ orbital in $^{24}$Mg
reduces the $B(E2:2^+_1\to 0^+_1)$ by 9\%. However,
low-lying spectra of a whole single-$\Lambda$ hypernucleus
have been difficult to calculate due to the unpaired $\Lambda$ particle.

In fact, a beyond mean-field calculation for
low-lying states of odd-mass nuclei based on modern energy density
functionals is a long-standing problem in nuclear physics.
One important reason for the difficulty is that the last unpaired
nucleon breaks some of the symmetries. Moreover, due to the
pairing correlation, many quasi-particle configurations are
close in energy and will be strongly mixed with each other.
Both of these facts complicate a calculation for
low-lying spectra of odd-mass nuclei at the beyond
mean-field level, although
some attempts have been made recently based on the
Skyrme energy density functional~\cite{Bally12}.

In this paper, we propose a novel microscopic particle-rotor model
for the low-lying states of single-$\Lambda$ hypernuclei.
The novel feature is that we combine the motion of $\Lambda$ particle
with the core nucleus states, which are described by the
state-of-the-art covariant density functional approach, that is,
the generator coordinate method (GCM) based on
the relativistic mean-field (RMF) approach supplemented with the
particle number and the angular momentum projections.

The particle-rotor model was firstly
proposed by Bohr and Mottelson \cite{BM75} (see also Ref. \cite{RS80}),
and has recently been applied also to
study the structure of odd-mass neutron-rich nuclei, such as
$^{11}$Be~\cite{PRM1,PRM2}, $^{15,17,19}$C~\cite{PRM3}, and
$^{31}$Ne~\cite{PRM4}.
In this model, the motion of a valence particle is coupled to
the rotational motion of a deformed core nucleus, which is usually
described by the rigid rotor model. The Pauli principle between
the valence nucleon and the nucleons in the core nucleus is treated
approximately.
In contrast to this conventional particle-rotor model,
in this paper we construct
low-lying states of the nuclear core
microscopically. That is, we superpose many quadrupole deformed RMF+BCS states,
after both the particle-number and
the angular-momentum projections are carried out~\cite{Yao10-14}.
A similar idea as the microscopic
particle-rotor model has recently been employed by Minomo {\it et al.}
in order to describe the structure of the one-neutron
halo nucleus $^{31}$Ne
with AMD \cite{MSK12}.
We apply this microscopic particle-rotor model to
hypernuclei, for which
the Pauli principle between the valence $\Lambda$ particle
and the nucleons in the core nucleus is absent.
We will demonstrate the applicability of this method
by studying the low-lying spectrum of $^{9}_{\Lambda}$Be.

{\em Formalism}.$-$
We describe a single-$\Lambda$ hypernucleus as a system in which
a $\Lambda$ hyperon interacts with nucleons inside a
nuclear core via a scalar and vector contact couplings.
The Lagrangian for the single-$\Lambda$ hypernucleus then reads,
\begin{equation}
 \label{lagrangian}
\mathcal{L}=\mathcal{L}_{\rm free}+\mathcal{L}_{\rm em}+\mathcal{L}^{NN}_{\rm int}+\mathcal{L}^{N\Lambda}_{\rm int},
\end{equation}
where $\mathcal{L}_{\rm free}$ is the free part of the Lagrangian
for the nucleons and the hyperon,
$\mathcal{L}_{\rm em}$ is the standard electromagnetic Lagrangian,
and $\mathcal{L}^{NN}_{\rm int}$ is
the effective strong interaction between
nucleons.
We employ a similar form for the $N\Lambda$ effective interaction term $\mathcal{L}^{N\Lambda}_{\rm int}$
as in Ref.~\cite{Tanimura2012},
and therefore, the vector and scalar $N\Lambda$ interaction
terms $\hat{V}_V^{N\Lambda}$ and $\hat{V}_S^{N\Lambda}$ are given by
\begin{eqnarray}
\label{interaction}
\displaystyle \hat{V}_V^{N\Lambda}(\vec{r}_{\Lambda},\vec{r}_{N})&=& \alpha_V^{N\Lambda} \delta(\vec{r}_{\Lambda}-\vec{r}_{N}) \\
\displaystyle \hat{V}_S^{N\Lambda}(\vec{r}_{\Lambda},\vec{r}_{N})&=& \alpha_S^{N\Lambda} \gamma^0_\Lambda \delta(\vec{r}_{\Lambda}-\vec{r}_{N})\gamma^0_N,
\end{eqnarray}
respectively.
For simplicity, the higher-order coupling terms and the
derivative terms in the $N\Lambda$ interaction are not
taken into account in the present study.

Based on the idea of particle-rotor model, we
construct
the wave function for single-$\Lambda$ hypernuclei
with an even-even nuclear core as
 \begin{equation}
 \label{wavefunction}
 \displaystyle \Psi_{IM}(\vec{r}_{\Lambda},\{\vec{r}_N\})
 =\sum_{j\ell I_c}  {\mathscr R}_{j\ell I_{c}}(r_{\Lambda}) {\mathscr F}^{IM}_{j\ell I_c}(\hat{\vec{r}}_{\Lambda}, \{\vec{r}_N\}),
\end{equation}
where
\begin{equation}
 {\mathscr F}^{IM}_{j\ell I_c}(\hat{\vec{r}}_{\Lambda}, \{\vec{r}_N\})
= [{\mathscr Y}_{j\ell}(\hat{\vec{r}}_{\Lambda})\otimes
\Phi_{I_c}(\{\vec{r}_N\})]^{(IM)}
\end{equation}
with $\vec{r}_{\Lambda}$ and $\vec{r}_N$
being the coordinates of the $\Lambda$ hyperon and the
nucleons, respectively.
In this equation,
$I$ is the total angular momentum and $M$ is its projection onto the
$z$-axis for the whole $\Lambda$ hypernucleus.
${\mathscr R}_{j\ell I_{c}}(r_{\Lambda})$
and ${\mathscr Y}_{j\ell}(\hat{\vec{r}}_{\Lambda})$ are
the four-component radial wave function and
the spin-angular wave function for the $\Lambda$ hyperon, respectively.

In the microscopic particle-rotor model,
the wave function for the nuclear core part, $\Phi_{I_cM_c}(\{\vec{r}_N\})$,
is given as a superposition of particle-number and angular-momentum
projected RMF+BCS states, $\vert \varphi(\beta)\rangle$, that is,
 \begin{equation}
 \vert \Phi_{I_cM_c}\rangle
 =\sum_\beta f_{I_c NZ}(\beta)
\hat P^{I_c}_{M_cK} \hat P^N\hat P^Z\vert \varphi(\beta)\rangle,
\label{GCM}
 \end{equation}
 where $\hat P^{I_c}_{M_cK}$, $\hat P^N$, $\hat P^Z$ are
the projection operators onto good numbers of angular momentum,
neutrons and protons, respectively. The  mean-field wave functions
$\vert \varphi(\beta)\rangle$ are a set of Slater determinants of
quasi-particle states with different quadrupole deformation $\beta$.
For simplicity, we consider only the axial deformation for the nuclear
core and thus the $K$ quantum number is zero in Eq. (\ref{GCM}).
The weight factor $f_{I_c NZ}(\beta)$
is determined
by solving the Hill-Wheeler-Griffin equation.
We call this scheme
a generator coordinate method (GCM) plus
particle-number (PN) and one-dimensional angular-momentum (1DAM)
projections, GCM+PN1DAMP.
See Ref. ~\cite{Yao10-14} for more
details on the GCM calculation for the nuclear core states.

Substituting
Eq. (\ref{wavefunction})
to the Dirac equation for the
whole hypernucleus, $H|\Psi_{IM}\rangle = E_I|\Psi_{IM}\rangle$,
where $H$ is the relativistic Hamiltonian corresponding to
Eq. (\ref{lagrangian}),
one can derive the coupled-channels equations for
${\mathscr R}_{j\ell I_{c}}(r_{\Lambda})$, in which
the coupling potentials are given in terms of
the transition densities.
From the solutions of those equations,
one can compute
the probability for the ($j\ell I_c$) component in the
total wave function, $\Psi_{IM}$, as
\begin{equation}
P_{j\ell I_c}=\int_0^\infty r_\Lambda^2 dr_\Lambda\,|{\mathscr R}_{j\ell I_{c}}(r_\Lambda)|^2.
\end{equation}
The reduced electric quadrupole ($E2$) transition strength
can be computed using the E2
operator, $\displaystyle\hat{Q}_{2\mu}=\sum_{i\in p} r_i^2 Y_{2\mu}(\hat{r}_i)$.
Notice that we use the bare charge in evaluating the $B(E2)$ strengths, that is,
+$e$ for protons and 0 for neutrons and a $\Lambda$ particle, since
our microscopic calculations are in the full configuration space.

\begin{figure}[]
\centering
\includegraphics[width=7.5cm]{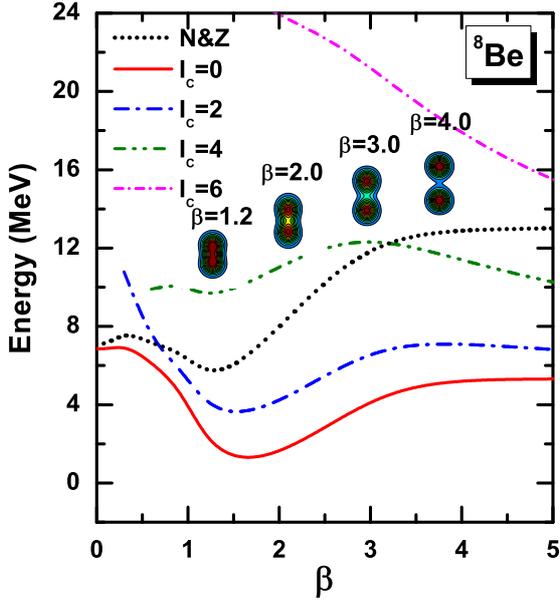}
\caption{The energy curve for $^8$Be
obtained in the mean-field approximation
with the particle number projection (the dotted line)
as a function of the intrinsic quadrupole deformation $\beta$.
The solid, the long dot-dashed, the dot-dot-dashed, and the
short dot-dashed lines
are projected energy curves for
$I_C=0,2,4,$ and 6, respectively.
The contour plots of the
intrinsic densities for $\beta=1.2, 2.0, 3.0$
and $4.0$ are also shown. }
\label{Be8RMF}
\end{figure}

{\em Results and discussions}.$-$ Let us now apply the microscopic particle-rotor model to $^9_\Lambda$Be
and discuss its low-lying spectrum. To this end,
we first carry out the GCM+PN1DAMP calculation
for the nuclear core, $^8$Be.
We generate the mean-field states $|\varphi(\beta)\rangle$
with constrained RMF+BCS calculations with quadrupole deformation.
The PC-F1 force~\cite{Buvenich02}, together with a density-independent
$\delta$ pairing force with a smooth cutoff factor~\cite{Krieger90},
is adopted. The pairing strengths are $V_n=-308$ and $V_p=-321$
MeV$\cdot$fm$^3$ for neutrons and protons, respectively.
More numerical details can be found in Ref.~\cite{Yao10-14}.
With the solutions of the GCM calculations,
we solve the coupled-channels equations
by expanding the radial
wave function ${\mathscr R}_{j\ell I_{c}}(r_\Lambda)$
on the basis of eigenfunctions of a spherical harmonic oscillator
with 18 major shells.
We take the same value for
the parameter $\alpha_S$ in the
$\Lambda$N interaction as in Ref.\cite{Tanimura2012} and vary the value of
$\alpha_V$ so that
the experimental $\Lambda$
binding energy of $^{9}_\Lambda$Be,
$B^{\rm{(exp.)}}_{\Lambda}(^{9}_{\Lambda}$Be)=6.71$\pm
0.04$ MeV~\cite{AGal1975},
is reproduced with the
microscopic particle-rotor model. The resultant values are
$\alpha_S^{N\Lambda}=-4.2377\times10^{-5}$ MeV$^{-2}$
and $\alpha_V^{N\Lambda}=1.2756\times10^{-5}$ MeV$^{-2}$.
The cut-off of the angular momentum for the core states ($I_c$)
is chosen to be 4, which gives well converged results for the
low-lying states of $^9_{\Lambda}$Be.
We include only the bound core states,
that is, the lowest energy state for each
value of $I_c$, even though all the possible states, including
continuum states, should be included in principle.

The dotted line in Fig. \ref{Be8RMF} shows the
energy curve for $^8$Be obtained in the mean-field approximation
with the particle number projection, as a
function of the
intrinsic quadrupole deformation $\beta$.
The energy curves with an additional projection
onto the angular momentum are also shown in the figure.
The density distributions in the mean-field approximation
are plotted for $\beta=1.2$, 2.0, 3.0, and 4.0, which
exhibit the two-$\alpha$ cluster structure.
After restoration of the rotational symmetry,
the energy minimum for $I_c=0$ is found at
$\beta=1.5$, while it is at $\beta=1.2$ in the mean-field approximation.
With the increase of the angular momentum $I_c$,
the energy curve
becomes softer and eventually the energy minimum
disappears at $I_c=6$.
It implies that the $6^+$ state in $^{8}$Be is unstable
against the 2$\alpha$ dissociation,
which is consistent with the experimental observation.

 \begin{figure*}[]
  \centering
 \includegraphics[width=16cm]{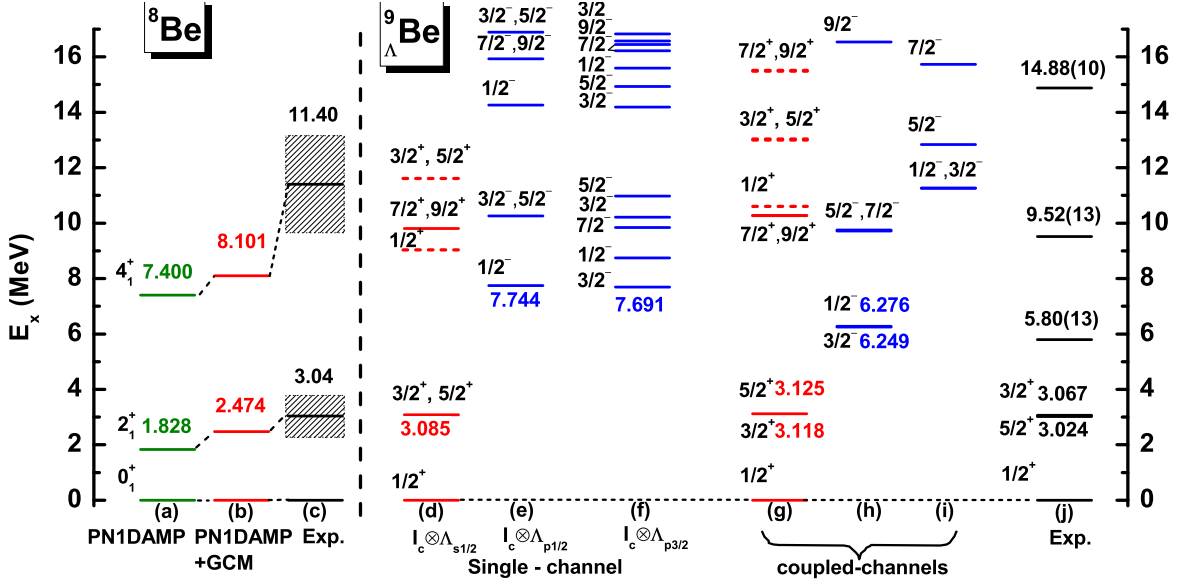}
 \caption{The low-energy excitation
spectra of $^{8}$Be (the columns (a), (b), and (c))
and $^{9}_\Lambda$Be (the columns (d) - (j)).
For $^{8}$Be, the full GCM+PN1DAMP calculations shown in the column (b)
are compared with those
without configuration mixing (PN1DAMP: the column (a))
for a single configulation at $\beta$=1.2.
The experimental data are taken from Ref.~\cite{Datar05}.
For $^{9}_\Lambda$Be, the columns (d), (e), and (f) show the results
of the single-channel calculations for the $\Lambda$ particle in
the $s_{1/2}$, $p_{1/2}$, and $p_{3/2}$ channels, respectively.
The columns (g), (h), and (i) show the results of the
coupled-channels equations,
which are compared with the experimental data
~\cite{Hashimoto06,Tamura05} shown in the column (j). }
  \label{spectra}
\end{figure*}

The full results of the GCM+PN1DAMP calculation for $^8$Be are shown
in
the column (b) in Fig.~\ref{spectra}. For a comparison,
the results with a single-configuration with $\beta=1.2$ and the
experimental data \cite{Datar05}
are also shown in the columns (a) and (c), respectively.
Notice that the former corresponds to the conventional particle-rotor
model, in which the nuclear shape, and thus, the deformation parameter is assumed to be identical
for each $I_c$ state.
Notice that in the full calculation the energy
minima appear at different deformations for different values of $I_c$ in Fig. 1.
One can see in Fig. 2 that the effect of
$I_c$ dependence of the deformation parameter as well as the
configuration mixing
increases the excitation energies of the $2^+_1$ and $4^+_1$ states
in $^{8}$Be, which are in closer agreement with the data~\cite{Datar05}.

We next discuss the spectrum of $^9_\Lambda$Be.
Before going to the full coupled-channels calculations, we first show the results of single-channel
calculations
in the columns (d), (e), and (f) in Fig. 2, where the
$\Lambda$ particle is in the $s_{1/2}, p_{1/2}$, and $p_{3/2}$ orbitals, respectively.
For the core nucleus states, we use the results of the GCM+PN1DAMP calculations
shown in the column (b) in Fig. 2.
For the $\Lambda$ particle in the $s_{1/2}$ orbit,
when it is coupled to the core excitation states of $2^+_1$ and $4^+_1$,
the degenerate ($3/2^+, 5/2^+$)
and ($7/2^+, 9/2^+$) doublet states
in $^{9}_\Lambda$Be are yielded, respectively.
We find that the excitation energies of these two doublet states
are slightly larger than those of the
corresponding excited states of the core nucleus.
This is caused by the fact that
the energy gain due to the $\Lambda$-N interaction is larger in the ground state as compared to that
in the other states.
For the $\Lambda$ particle in the $p_{1/2}$ and $p_{3/2}$ orbitals,
one obtains the lowest negative parity $1/2^-$ and $3/2^-$
states in $^{9}_\Lambda$Be.
The $1/2^-$ state is higher than the $3/2^-$ state by 0.03 MeV,
which reflects the size of spin-orbit splitting
in $p_\Lambda$ state of $^{9}_\Lambda$Be.
The $1/2_1^-$, $7/2_1^-$, $3/2_2^-$, and $5/2_1^-$ states
around 10 MeV in the column (f) in Fig. 2 are resulted from the
2$^+\otimes p_{3/2}$ configuration. The order of these states
can be understood in terms of the reorientation effect, that is, the diagonal
component for the quardupole term in the coupling potential
in the coupled-channels equations.
On the other hand,
for the
2$^+\otimes p_{1/2}$ configuration, in which the $\Lambda$ particle in the $p_{1/2}$ orbital coupled to the
2$^+$ state of the core nucleus,
the quadrupole term does not contribute, and the 3/2$^-$ and 5/2$^-$
states are degenerate in energy in the column (e) in Fig. 2.
A more detailed discussion on this characteristic appearance of the multiples will
be given in the forthcoming publication \cite{MHYM14}.

The low-energy spectrum of $^{9}_{\Lambda}$Be, obtained by
mixing these single-channel configurations with
the coupled-channels method,
is shown
in the column (g), (h), and (i) in Fig. 2. The low-lying states are
categorized into
 three rotational bands, whose structures are confirmed by the calculated $B(E2)$ relations.
It is remarkable that the present calculation reconfirms such an
interesting prediction of the cluster model
that the strong coupling of a hyperon to the collective rotation is realized when
the $\Lambda$ is in the $p$-orbit \cite{Motoba83}.
Among these rotational bands,
the column (h) corresponds to what they called genuine
hypernuclear states ~\cite{Motoba83},
which are also referred to as the
supersymmetric states having the SU(3) symmetry $(\lambda\mu)=(50)$
of $s^4p^5$ shell-model configuration \cite{DG76}.
These states do not have corresponding states in the
ordinary nucleus, $^9$Be, because of the Pauli principle of the valence neutron.
The calculated spectrum
is compared with
the available data~\cite{Hashimoto06,Tamura05}
shown in the column (j) in Fig.~\ref{spectra}. One can see that
a good agreement with the data is obtained
with our calculations.
According to our calculations, the measured state with excitation
energy of 5.80(13) MeV is actually a mixture of
two negative-parity states
with $J^\pi=3/2^-$ and $1/2^-$.
The state with excitation energy of
9.52(13) MeV, on the other hand,
would be a mixture of $J^\pi=9/2^+, 7/2^+,
7/2^-, 5/2^-$, and $1/2^+$ states,
which are close in energy.

\begin{table*}[]
\centering
\tabcolsep=6pt
 \caption{The probability $P_{jlI_c}$ of the dominant components
in the wave function for low-lying states of $^{9}_{\Lambda}$Be obtained
by the microscopic particle-rotor model.
Only those components which have $P_{jlI_c}$ larger than 0.1 are shown.
$E$ is the energy of each state obtained by solving the coupled-channels
equations, while $E_{\rm 1ch}^{(0)}$ is the unperturbed energy obtained
with the single-channel calculations. The energies are listed in units of
MeV. }
 \begin{tabular}{ccccc|ccccc}
  \hline\hline
     $I^\pi$      &  $E$       &$(l~j)\otimes I_c$    &  $P_{jlI_c}$       &
$E^{(0)}_{\rm 1ch}$         &  $I^\pi$      &  $E$        & $(l~j)\otimes I_c$         &  $P_{jlI_c}$        &  $E^{(0)}_{\rm 1ch}$                      \\
  \hline
     $1/2^+_1 $   &  $0.000 $   &  $s_{1/2}\otimes0^+$ &  $0.928 $  &  $0.000 $   &       $1/2^+_2 $   &  $10.603$   &  $s_{1/2}\otimes0^+$ &  $0.995 $  &  $9.035$                 \\
     $        $   &  $      $   &  $                 $ &  $      $  &  $      $   &        $        $   &  $      $   &  $                 $ &  $      $ &  $      $                     \\
     $3/2^+_1 $   &  $3.118$    &  $s_{1/2}\otimes2^+$ &  $0.919 $  &  $3.085 $   &       $3/2^+_2 $   &  $13.034$   &  $d_{3/2}\otimes0^+$ &  $0.841 $  &  $11.467 $                    \\
     $        $   &  $      $   &  $                 $ &  $      $  &  $      $   &       $        $   &  $      $   &  $s_{1/2}\otimes2^+$ &  $0.131 $  &  $11.610$                 \\
     $5/2^+_1 $   &  $3.125 $   &  $s_{1/2}\otimes2^+$ &  $0.919 $  &  $3.085 $   &       $5/2^+_2 $   &  $12.999$   &  $d_{5/2}\otimes0^+$ &  $0.845 $  &  $ 11.450 $                    \\
     $        $   &  $      $   &  $                 $ &  $      $  &  $      $   &       $        $   &  $      $   &  $s_{1/2}\otimes2^+$ &  $0.125 $  &   $11.610$                 \\
     $7/2^+_1 $   &  $10.267$   &  $s_{1/2}\otimes4^+$ &  $0.894$   &  $9.807 $   &       $7/2^+_2 $   &  $15.510$   &  $d_{3/2}\otimes2^+$ &  $0.833 $  &  $ 13.804$                     \\
     $        $   &  $      $   &  $                 $ &  $      $  &  $      $   &       $        $   &  $      $   &  $d_{5/2}\otimes2^+$ &  $0.112 $  &  $14.124$                      \\
     $9/2^+_1 $   &  $10.281$   &  $s_{1/2}\otimes4^+$ &  $0.894 $  &  $9.807 $   &       $9/2^+_2 $   &  $15.483$   &  $d_{5/2}\otimes2^+$ &  $0.943 $  &  $13.734$                      \\
     $        $   &  $      $   &  $                 $ &  $      $  &  $      $   &       $        $   &  $      $   &  $                 $ &  $      $  &  $      $                     \\
     $1/2^-_1   $ &  $6.276$   &  $p_{1/2}\otimes0^+$ &  $0.516$  &  $  7.744 $   &       $1/2^-_2   $ &  $11.271 $  &  $p_{1/2}\otimes0^+$ &  $0.575 $  &  $7.744 $               \\
     $          $ &  $      $   &  $p_{3/2}\otimes2^+$ &  $0.445 $  &  $  8.741 $ &       $          $ &  $       $  &  $p_{3/2}\otimes2^+$ &  $0.419$   &  $8.741 $               \\
     $3/2^-_1   $ &  $6.249$   &  $p_{3/2}\otimes0^+$ &  $0.524 $  &  $  7.691 $  &       $3/2^-_2   $ &  $11.258 $  &  $p_{3/2}\otimes0^+$ &  $0.568 $  &  $7.691 $               \\
     $          $ &  $      $   &  $p_{3/2}\otimes2^+$ &  $0.220 $  &  $ 10.212 $ &       $          $ &  $       $  &  $p_{3/2}\otimes2^+$ &  $0.220 $  &  $10.212$                \\
     $          $ &  $      $   &  $p_{1/2}\otimes2^+$ &  $0.217 $  &  $ 10.259 $ &       $          $ &  $       $  &  $p_{1/2}\otimes2^+$ &  $ 0.206$  &  $10.259 $              \\
     $5/2^-_1   $ &  $9.756$   &  $p_{1/2}\otimes2^+$ &  $0.625 $  &  $  10.259 $ &       $5/2^-_2   $ &  $12.835 $  &  $p_{3/2}\otimes2^+$ &  $0.769 $  &  $10.978 $               \\
     $          $ &  $      $   &  $p_{3/2}\otimes2^+$ &  $0.186 $  &  $10.978$   &       $          $ &  $       $  &  $p_{1/2}\otimes2^+$ &  $0.226 $  &  $10.259$               \\
     $          $ &  $      $   &  $p_{3/2}\otimes4^+$ &  $0.150 $  &  $14.935$   &       $7/2^-_2   $ &  $15.729 $  &  $p_{3/2}\otimes2^+$ &  $ 0.603$  &  $9.842 $              \\
     $7/2^-_1   $ &  $9.717 $   &  $p_{3/2}\otimes2^+$ &  $0.813 $  &  $~~9.842 $ &       $          $ &  $       $  &  $p_{1/2}\otimes4^+$ &  $0.216 $  &  $15.926$               \\
     $        $   &  $      $   &  $p_{1/2}\otimes4^+$ &  $0.100 $  &  $15.926$   &       $          $ &  $       $  &  $p_{3/2}\otimes4^+$ &  $ 0.157$  &  $ 16.215$        \\
\hline \hline
 \end{tabular}
    \label{tab1}
   \end{table*}

\begin{table}[]
\tabcolsep=2pt
\caption{The calculated $E2$ transition strengths (in units of $e^2$ fm$^4$)
for low-lying states of $^{8}$Be and $^{9}_{\Lambda}$Be.
c$B(E2)$ is defined by Eq.(\ref{cBE2}), where $L$ is the total orbital
angular momentum.
}
\label{BE2}
\begin{center}
 \begin{tabular}{cc|cccc}\hline\hline
 \multicolumn{2}{c|}{$^{8}$Be}& \multicolumn{4}{c}{$^{9}_{\Lambda}$Be }  \\ \hline
 $I^\pi_i \to I^\pi_f$       & $B(E2)$      &  $I^\pi_i \to I^\pi_f$
&($L^\pi_i \to L^\pi_f$) & $B(E2)$ & cB(E2)  \\ \hline
 $2^+_1\rightarrow 0^+_1$    & $24.99 $   & $3/2^+_1\rightarrow 1/2^+_1$& ($2^+\to 0^+$)& $22.55$ & $22.55$ \\
 $                      $    & $      $   & $5/2^+_1\rightarrow 1/2^+_1$& ($2^+\to0^+$) & $22.57$ & $22.57$ \\
 $4^+_1\rightarrow 2^+_1$    & $47.28 $   & $7/2^+_1\rightarrow 3/2^+_1$& (4$^+\to2^+$) & $37.43$ & 41.58\\
 $                      $    & $      $   & $9/2^+_1\rightarrow 5/2^+_1$& (4$^+\to2^+$) & $41.55$ & $41.55$ \\
 $                      $    & $      $   & $7/2^+_1\rightarrow 5/2^+_1$& (4$^+\to2^+$) & $4.152$ &41.52 \\
 $                      $    & $      $   & $5/2^-_1\rightarrow 1/2^-_1$& $(3^-\to1^-$) & $13.14$ & $16.90$ \\
 $                      $    & $      $   & $7/2^-_1\rightarrow 3/2^-_1$& (3$^-\to1^-$) & $17.15$ & $17.15$ \\ \hline
\end{tabular}
\end{center}
\end{table}

Table~\ref{tab1} lists the values
of the probability of the dominant components
(with $P_{j\ell I_c}\geq 0.10$) for a few low-lying states of
$^{9}_{\Lambda}$Be. The unperturbed energies, $E^{(0)}_{\rm 1ch}$, obtained
by the single-channel calculations are also shown for each component.
One can see that the positive-parity states in the ground
state rotational band
are almost pure $I^+_c\otimes \Lambda_{s 1/2}$ states,
while there are appreciable configuration mixings for
the negative-parity states as well as the positive-parity
states in the excited band.
For instance,
for the first negative-parity state, $1/2_1^-$,
there is a strong mixing
between the 0$^+\otimes \Lambda_{p_{1/2}}$ and the 2$^+\otimes \Lambda_{p_{3/2}}$
configurations with almost equal weights.
These large mixtures of collective core wave functions manifest
the strong coupling mediated by the $p$-state hyperon.
This is caused by the fact that
the unperturbed energies, $E_{\rm 1ch}^{(0)}$,
are similar to each other for these configurations due to the reorientation
effect.
It is interesting to notice that
the values of $P_{j\ell I_c}$ obtained in the present calculations
are similar to those with the cluster model calculations shown in Fig. 2
of Ref. \cite{Motoba83}.
We also remark that in the second
positive parity states ($I_2^+$) the $\Lambda_d$ state is admixed appreciably, while in the second negative
parity states ($I_2^-$) the wave functions have the "out-of-phase" nature
in comparison with the corresponding first negative parity states ($I_1^-$).

Table~\ref{BE2} shows the calculated $E2$ transition strengths
for low-lying states of $^{8}$Be and $^{9}_{\Lambda}$Be.
As has been pointed out in Ref. \cite{Motoba83},
each state in $^9_\Lambda$Be can be classified
in terms of the total orbital angular momentum $L$, which
couples to the spin 1/2 of the $\Lambda$ particle to
form the total angular momentum $I$.
In order to remove the trivial
factor due to the angular momentum coupling for
spin 1/2 and
see more clearly the impurity effect of $\Lambda$ particle on
nuclear collectivity, we follow Ref. \cite{Motoba83} and
compute
the c$B(E2)$ value defined as,
\begin{eqnarray}
&& cB(E2: L_i\rightarrow L_f) \nonumber \\
&& \equiv \hat{L_i}^{-2}\hat{I_f}^{-2} \left\{ \begin{matrix}
L_f & I_f     &  1/2 \\
I_i     & L_i &  2
 \end{matrix} \right\}^{-2}
 B(E2:I_i\rightarrow I_f),
\label{cBE2}
\end{eqnarray}
where $\hat{I}\equiv \sqrt{2I+1}$.
Notice that this quantity is
more general than
the one introduced in
Ref.~\cite{Isaka11}, since the weak coupling does not have to be assumed,
although
both formulas are equivalent for
a single-channel configuration of $|\Lambda_{s_{1/2}}\otimes \Phi_{I_c}\rangle$.
The impurity effect of $\Lambda$ particle
on $^{8}$Be can be discussed by comparing the $B(E2)$ values in $^{8}$Be
and the c$B(E2)$ values in $^{9}_\Lambda$Be in Table 2.
One can see that the $E2$ strengths are slightly decreased
due to the shrinkage effect of $\Lambda$ hyperon.

The degree of reduction in $B(E2)$ with
the present model
is much smaller than the results
of cluster model calculations \cite{Motoba83,HK11}.
Also, the energy of the 2$^+$ state increases significantly by adding a
$\Lambda$ particle, whereas the experimental data indicate that the energy
shift is negligibly small \cite{HK11} (see Fig. 2).
These would be due to the effects of higher members of the core
excited states, which are not included in
present calculations.
However, with the current implementation of GCM,
we have a limitation in constructing the wave functions for the
non-resonant core states (that is, the second and the third
0$^+$, 2$^+$, and 4$^+$ states), which extend
up to large values of deformation
parameter $\beta$. For this reason, we did not obtain reasonable
results for the hypernuclear states above 5/2$^+_1$
when we included the higher members of the core states.

{\em Summary}.$-$
In summary, we have proposed a novel
method for a low-lying spectrum of hypernuclei based on a
mean-field type approach.
Whereas the pure mean-field approximation does not yield
a spectrum due to the broken rotational symmetry,
we employed a beyond relativistic mean-field approach by
carrying out the angular momentum and the particle number projections
as well as the configuration mixing with the
generator coordinate method.
In this novel method,
the beyond mean-field approach is applied to
low-lying states of
the core nucleus, to which
the $\Lambda$ hyperon couples in
the wave function of hypernuclei,
and thus we call
it the microscopic particle-rotor model.
We emphasize that this is the
first calculation for a spectrum of hypernuclei
based on a density functional approach.
By applying the microscopic particle-rotor model to
$^9_\Lambda$Be,
a reasonable agreement with the experimental data of
low-spin spectrum has been achieved without introducing
any adjustable parameters, except for the $N\Lambda$ coupling strengths,
which were determined to reproduce the $\Lambda$ binding energy.

In this paper, we have assumed the axial deformation for the core nucleus
$^8$Be. An obvious extension of our method is to take into account
the triaxial deformation of the core nucleus.
One interesting candidate for this is
$^{25}_{\Lambda}$Mg, for which
the triaxial degree of freedom has been shown to be important
in the core nucleus $^{24}$Mg\cite{Win11,Yao10-14,RB68}.
Another point which we would like to make is that
our method is not restricted to the rotational motion of a core nucleus,
but the vibrational motion can also be treated on the equal footing
using the generator coordinate method.
It will be interesting to
apply systematically
the present method to many hypernuclei and to study
a transition in low-lying spectrum from a vibrational to a
rotational characters.
An application of our method to ordinary nuclei with an odd number of nucleons is
another interesting problem, although a treatment of the Pauli principle would make
it more complicated as compared to hypernuclei studied in this paper.

%
\section*{Acknowledgments}
This work was supported in part by the Tohoku University Focused Research Project ``Understanding the origins for matters in universe",
JSPS KAKENHI Grant Number 2640263, the National Natural Science Foundation of China
under Grant Nos. 11305134, 11105111, and the Fundamental Research Funds for the Central
Universities (XDJK2010B007 and XDJK2013C028).

%
%


\begin{thebibliography}{99}{}
 \bibitem{Hashimoto06} O. Hashimoto and H. Tamura, Part. Nucl. Phys. \textbf{57} (2006) 564.
 \bibitem{Tamura09} H. Tamura, Int. J. Mod. Phys. A \textbf{24} (2009) 2101.

 \bibitem{Motoba83}T. Motoba, H. Band\={o}, and K. Ikeda, Prog. Theor. Phys. \textbf{70} (1983) 189.

\bibitem{Motoba85} T. Motoba, H. Band\={o}, K. Ikeda, and T. Yamada, Prog. Theor. Phys. Suppl. \textbf{81} (1985) 42.

 \bibitem{Hiyama99} E. Hiyama, M. Kamimura, K. Miyazaki, and T. Motoba, Phys. Rev. C \textbf{59} (1999) 2351.

 \bibitem{Tanida01} K. Tanida {\it et al.}, Phys. Rev. Lett. \textbf{86} (2001) 1982.

 \bibitem{Bando90}  H. Bando, T. Motoba and J. \v{Z}ofka, Int. J. Mod. Phys. \textbf{A 5} (1990) 4021.
 \bibitem{Hiyama03} E. Hiyama, Y. Kino, and M. Kamimura, Prog. Part. Nucl. Phys. \textbf{51} (2003)  223.

 \bibitem{Dalitz78} R. H. Dalitz and A. Gal, Ann. Phys. (N.Y.) \textbf{116} (1978) 167.

\bibitem{Gal71}A. Gal, J.M. Soper, and R.H. Dalitz, Ann. Phys. (N.Y.) {\bf 63} (1971) 53.

\bibitem{Millener}D.J. Millener, Nucl. Phys. {\bf A804} (2008) 84; {\bf A914} (2013) 109.

\bibitem{abinitio}R. Wirth, D. Gazda, P. Navratil, A. Calci, J. Langhammer, and R. Roth, arXiv:1403.3067 [nucl-th] (2014).

\bibitem{Isaka11} M. Isaka, M. Kimura, A. Dot\'e and A. Ohnishi,
                  Phys. Rev. C \textbf{83} (2011) 044323; Phys. Rev. C \textbf{83} (2011) 054304.

\bibitem{Bender03}M. Bender, P.-H. Heenen, and P.-G. Reinhard, Rev. Mod. Phys. {\bf 75} (2003) 121.

 \bibitem{Zhou07} X. R. Zhou, H.-J. Schulze, H. Sagawa, C. X. Wu, and E.-G. Zhao, Phys. Rev. C \textbf{76} (2007) 034312.
 \bibitem{Win08} M. T. Win and K. Hagino, Phys. Rev. C \textbf{78} (2008) 054311.
 \bibitem{Schulze10} H.-J. Schulze, M. T. Win, K. Hagino, and H. S. Sagawa, Prog. Theo. Phys. \textbf{123} (2010) 569.
 \bibitem{Win11} Myaing Thi Win, K. Hagino, and T. Koike, Phys. Rev. C \textbf{83} (2011) 014301.
 \bibitem{Lu11} B.-N. Lu, E.-G. Zhao, and S.-G. Zhou, Phys. Rev. C \textbf{84} (2011) 014328.
 \bibitem{Li13} A. Li, E. Hiyama, X.-R. Zhou, and H. Sagawa, Phys. Rev. C \textbf{87} (2013) 014333.
 \bibitem{Lu14} B.-N. Lu, E. Hiyama, H. Sagawa, and S.-G. Zhou, Phys. Rev. C \textbf{89} (2014) 044307.

\bibitem{Yao11NPA} J. M. Yao, Z. P. Li, K. Hagino, M. Thi Win, Y. Zhang, and J. Meng, Nucl. Phys. \textbf{A868-869} (2011) 12.

\bibitem{Bally12} B. Bally, B. Avez, M. Bender, and P.-H. Heenen, Int. J. Mod. Phys. E \textbf{21}  (2012) 1250026.

\bibitem{BM75}A. Bohr and B.R. Mottelson, {\it Nuclear Structure Vol. II}
(Benjamin, Reading, MA, 1975).

\bibitem{RS80} P. Ring and P. Schuck,
{\it The Nuclear Many Body Problem}
(Springer-Verlag, New York, 1980).

\bibitem{PRM1} H. Esbensen, B. A. Brown, and H.Sagawa,
Phys. Rev. C \textbf{51} (1995) 1274.
\bibitem{PRM2} F. M. Nunes, I. J. Thompson, and R. C. Johnson, Nucl. Phys. A \textbf{596} (1996) 171.
\bibitem{PRM3} T. Tarutina and M. S. Hussein, Phys. Rev. C \textbf{70} (2004) 034603.
\bibitem{PRM4}Y.Urata, K.Hagino, and H.Sagawa, Phys. Rev. C \textbf{83} (2011) 041303(R).

\bibitem{Yao10-14} J. M. Yao, J. Meng, P. Ring, and D. Vretenar,
Phys. Rev. C \textbf{81} (2010) 044311;
J. M. Yao, H. Mei, H. Chen, J. Meng, P. Ring, and
D. Vretenar, Phys. Rev. C \textbf{83} (2011) 014308;
J. M. Yao, H. Mei, Z. P. Li,
Phys. Lett. B \textbf{723}  (2013) 459;
J. M. Yao, K. Hagino, Z. P. Li, J. Meng, and P. Ring,
Phys. Rev. C  \textbf{89} (2014) 054306.


\bibitem{MSK12}K. Minomo, T. Sumi, M. Kimura, K. Ogata, Y.R. Shimizu,
and M. Yahiro, Phys. Rev. Lett. {\bf 108}  (2012) 052503.

\bibitem{Tanimura2012}Y. Tanimura and K. Hagino,
                      Phys. Rev. C \textbf{85} (2012) 014306.

 \bibitem{Buvenich02} T. Burvenich, D. G. Madland, J. A. Maruhn, and P. G. Reinhard, Phys. Rev. C \textbf{65} (2002) 044308.

\bibitem{Krieger90} S. J. Krieger, P. Bonche, H. Flocard, P. Quentin, and M. S. Weiss,
                  Nucl. Phys. A \textbf{517} (1990) 275.

\bibitem{AGal1975}A. Gal, Advances in Nucl. Phys. {\bf 8} (1975)  1.

  \bibitem{Datar05} V. M. Datar, Suresh Kumar, D. R. Chakrabarty, V. Nanal, E. T. Mirgule, A. Mitra, and H. H. Oza, Phys. Rev. Lett. \textbf{94} (2005)  122502.

\bibitem{MHYM14}H. Mei, K. Hagino, J.M. Yao, and T. Motoba, to be published.

\bibitem{DG76}R.H. Dalitz and A. Gal, Phys. Rev. Lett. {\bf 36} (1976) 362.

\bibitem{Tamura05} H. Tamura, S. Ajimura, H. Akikawa {\it et al.}, Nucl. Phys. A \textbf{754} (2005) 58c.

\bibitem{HK11}K. Hagino and T. Koike, Phys. Rev. C{\bf 84} (2011) 064325.

\bibitem{RB68}S.W. Robinson and R.D. Bent, Phys. Rev. {\bf 168} (1968) 1266.


\end{thebibliography}
\end{document}